\documentclass{elsart}

 \usepackage{graphicx}
  \include{epsfig}
\journal{Physica A}
\begin{document}
\begin{frontmatter}


\title{Stylized Facts Generated Through Cellular Automata Models. Case
of Study: The Game of Life}
\author{H.F. Coronel-Brizio},
\author{A.R. Hern\'andez-Montoya\corauthref{cor}},
\ead{alhernandez@uv.mx}
\ead[url]{www.uv.mx/alhernandez}
\author{M.E. Rodr\'{\i}guez-Achach} and \author{G.A. Stevens-Ram\'{\i}rez}

\corauth[cor]{Corresponding author: Maestr\'{\i}a en Inteligencia Artificial. Sebasti\'an Camacho 5, Xalapa Veracruz 91000, M\'exico. Tel/Fax: 52-228-8172957/8172855.}

\address{Facultad de F\'{\i}sica e Inteligencia Artificial.
Universidad Veracruzana, Apdo. Postal 475. Xalapa, Veracruz. M\'{e}xico}

\begin{abstract}
\noindent
In the present work, a geometrical method to generate a two
dimensional random walk by means of a bidimensional Cellular Automaton
is presented. We illustrate it by means of Conway's Game of Life with
periodical borders, with a  large lattice of $3000 \times 3000$
cells. The obtained random walk is of character anomalous, and its
projection to a one dimensional random walk is analyzed, showing that it
presents some statistical properties similar to the so-called
stylized facts observed in financial time series. We consider that
the procedure  presented here is important not only because of its simplicity, 
but also because it could help us to understand and shed light on the
stylized facts formation mechanism.
\end{abstract}

\end{frontmatter}


\vspace*{-1.5cm}
 \section{Introduction}
\vspace*{-.9cm}
\noindent
Recent methodologies from Physics have been successfully 
applied to the study of Economy and Financial Markets Complexity. 
This new area of research is called Econophysics. The econophysics 
community is currently doing active research in topics such as the study 
of the distributional properties of the variations of the Stock Market, 
Network Analysis of Economical Phenomena, Financial Crashes, Wealth Distributions, study of Financial Markets under a Microscopic point of view, etc. 
About the latter, it involves techniques named Microscopic Simulation 
(MS) \cite{MS}, that consists of the study of a complex system by 
individually following on a computer each agent of the system and its 
interactions with the other agents, and simulating the overall system 
evolution. This technique is becoming very promising and useful. 
Following this direction, very interesting models that reproduce the 
statistical properties of financial Markets, named by the economics 
community ``Stylized Facts" \cite{Fama,Cont}, have been proposed 
\cite{MG,Lux}. Also in a very related way, a great amount of work 
has been carried out by constructing artificial stock markets by means 
of Cellular Automata models \cite{qiu,Thomas,zhou}.\\
\noindent
Cellular Automata (CA) are spacetime-like discrete deterministic dynamical 
systems whose behavior is defined completely in terms of local interactions. 
Cellular automata were introduced by John Von Neumann who intended  to 
understand the biological mechanisms of self-reproduction \cite{vonneuman} 
and have now become an important object of study because of their intrinsic 
mathematical interest as well as their success as a tool  to model complex 
phenomena in physical, chemical, economical and biological systems, design of parallel computing architectures, traffic models, programming 
environments, etc \cite{elsewhere}.\\
\noindent
Cellular automata became very popular at the beginning of the 70's thanks 
to an article written by Martin Gardner and published in  \textit
{Scientific American} \cite{gardner},  about  the cellular automaton
called \textit {The Game of Life} (GOL) or just \textit{Life}. This
cellular automaton  was invented by the mathematician John H. Conway
at the end of 60's and since, it displayed a very rich and interesting
complex behavior; very soon it became the favorite game of the --at that time
incipient-- community of computer fans. 

\vspace*{-.9cm}
\subsection {The Game of Life}
\vspace*{-.9cm}
\noindent 
Life is a class IV bi-dimensional (shows complex behavior) totalitarian cellular automaton \cite{wolf1}. The updating rule that determines the Game of Life 
evolution is applied on a Moore neighborhood as follows: a) a dead
cell surrounded by exactly three living cells is born again. b) a live cell 
will die if either it has less than two or more than three living neighbors.
\noindent
This simple rule produces a very rich and complex behavior, generating 
self-organized structures and also producing very important and interesting 
emergent properties (formation of self-replicating structures, Universal 
Computation, etc). It is for this reason that we have chosen Life to generate 
a bi-dimensional random walk and then a time series with expected complex 
characteristics.
\noindent
In section \ref{walk} we explain how, by using a geometrical procedure applied 
to Life, we can generate a time series $r_i$, analyzed in this work, that
shows statistical properties very similar to those of Financial Time Series, 
known in the economic community as stylized facts. In section \ref{stylized} we 
show the numerical results corresponding to these analyses.

\vspace*{-1.1cm}
\section{Generating a random walk by the Game of Life. Defining our observable}
\label{walk}
\begin{figure}[htb!]
\begin{center}
\resizebox{0.75\textwidth}{!}{%
\includegraphics{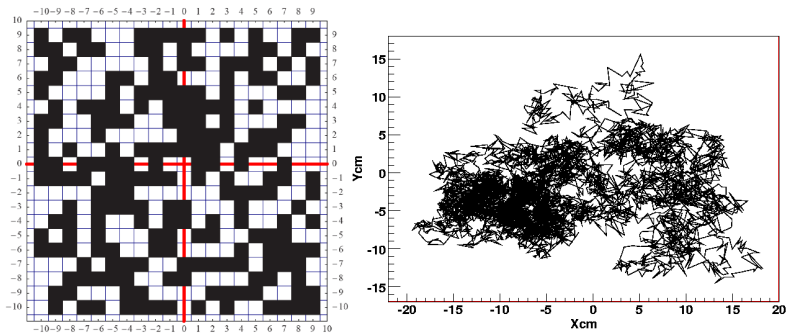}}
\caption{Left panel: Coordinate system used to calculate the vector $\vec{R}(i)_{CM}$ for 
each state of the GOL CA (only for illustration purposes). Right panel: Random Walk 
generated when $\vec{R}(i)_{CM}$ evolves in time.}
\label{rw}
\end{center}
\end{figure}

\vspace*{-.9cm}
\noindent
\noindent
Due to the need of dealing with a finite size $N \times N$ of the CA 
lattice, in our work we have chosen to program our CA using periodical 
boundary conditions.  After centering a Cartesian coordinate system $XY$ in the 
lattice of the CA as shown in figure \ref{rw}a), in order to obtain the one-dimensional 
observable analyzed in this paper, we construct the position vector $\vec{R}(i)_{CM}$ 
of a point following a bidimensional random walk\footnote{This random walk is of 
character anomalous, but this is demonstrated in a forthcoming paper.} obtained 
for each time step of the $M$ generated points, $i= 1,2,3,..,M$,  as follows:
\vspace*{-.6cm}
\begin{center}
\begin{equation}
\label{eq:mass_center_gral}
\vec{R}(i)_{CM} := ( X(i)_{CM}, Y(i)_{CM}) = 
\frac{1}{N}\sum_{k=1}^{N}\sum_{l=1}^{N}C_{kl}(i)(x_{k}(i),y_{l}(i)), 
\end{equation}
\end{center}
\vspace*{-.6cm}
\noindent
where $N \times N$, is the total number of cells of the CA, $C_{kl}(i)$ denotes the
state (1 or 0) of the cell in the  coordinates $(x_{k}(i),y_{l}(i))$ with respect to the coordinate system $XY$ at time step $i$. 
\noindent
So, while the CA evolves, this vector\footnote{Any resemblance with
  the Center of Mass (CM) definition of a mass distribution is true,  $\vec{R}(i)_{CM}$ describes how 
the "center of mass" of living cells distribution evolves.} is calculated for 
each time step $i=1,2,3,..,M$. Figure \ref{rw}b) shows $10000$ time steps of the evolution 
of the vector $\vec{R}(i)_{CM}$. Our observable will be the distance to the origin 
$r_i$ of $\vec{R}(i)_{CM}$, i.e. $r_i :=  \sqrt{X(i)_{CM}^2+Y(i)_{CM}^2}$, $i=1,2,3,..,M$.  
In order to analyze the one-dimensional time series generated by $r_i$ as time goes on, 
we construct the returns or logarithmic differences $S_i$, of the time series $r_i$ as 
usual: $S_i: = \log r_{i+1} - \log r_i$, for $i = 1,2,3,..,M$. 

\vspace*{-.7cm}
\subsection{Data Sample}
\vspace*{-.9cm}
\noindent
In order to have certain control of GOL initial states, configurations 
of $20\%$, $40\%$ and  $60\%$ of living cells were initially set up at random 
on lattices of size $3000\times 3000$. For each initial density, 20 random walks 
of 20000 steps were generated, giving us a total $12 00000$ steps to analyze.
Considering certain characteristics such as the finite size of the GOL lattice,
or a particular initial configuration, GOL generated fluctuations tend to die out 
after an undetermined number of time steps. Because of this we have applied a cut off 
to  $r_i$, $i=1,2,3,..,M$, considering 
its values such as $M < 3000$. Figures of next section will show the reason of 
this data selection more clearly.\\

\vspace*{-.9cm}
\section{Numerical Results: Statistical Properties of $r_i$ returns}
\label{stylized}
\begin{figure}[htb!]
\begin{center}
\resizebox{0.6\textwidth}{!}{%
\includegraphics{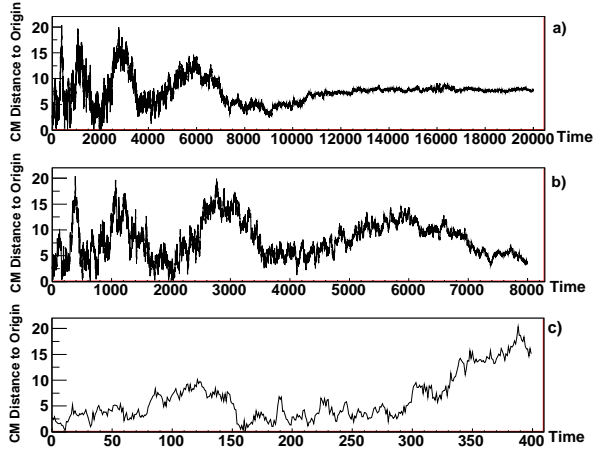}}
\caption{Time evolution of our observable, $\vec{R}(i)_{CM}$ distance to Origin for
  time ranges of a) $20000$  b) $8000$  and c) $400$ time steps.}
\label{serie}
\end{center}
\end{figure}

\begin{figure}[htb!]
\begin{center}
\resizebox{0.6\textwidth}{!}{%
\includegraphics{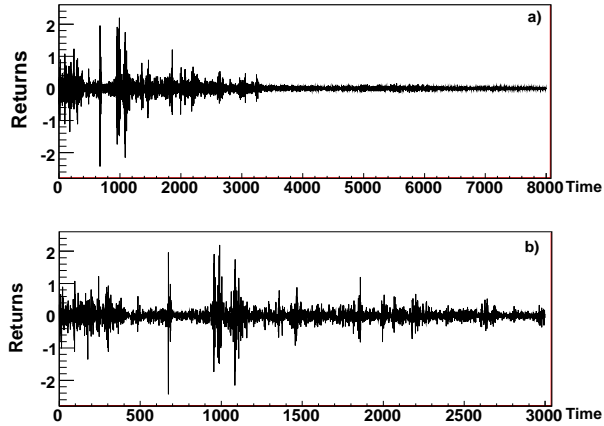}}
\caption{Returns evolution. In a) is clear the effect of the CA fluctuations dying out. b) Zooming in the upper figure.  We can appreciate the clustering volatility effect. }
\label{cluster}
\end{center}
\end{figure}

\vspace*{-.9cm}
\noindent
In this section we will show that the time series $r_i$, $i=1,2,3,..,M$ has statistical 
properties that resemble the stylized facts of financial time series. Figure \ref{serie} 
shows a plot of $r_i$ for three different time ranges, while figure \ref{cluster} shows 
the graphic of returns vs. time for two different time ranges. Both figures were obtained from one of the typical generated random walks. 

\vspace*{-.7cm}
\subsection{Returns Distributions}
\begin{figure}[htb!]
\begin{center}
\resizebox{0.6\textwidth}{!}{%
\includegraphics{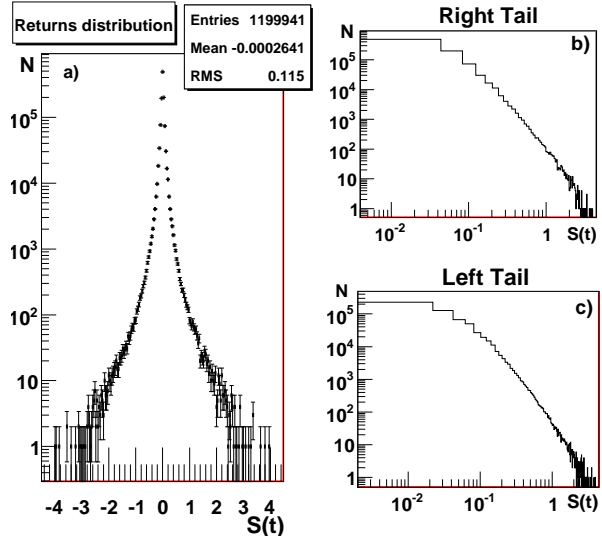}}
\caption{Returns Distribution. a) Full Sample, b) Returns Distribution Log-Log Plot for right tail; c) Returns Distribution Log-Log Plot for left tail.}
\label{Returns}
\end{center}
\end{figure}

\begin{figure}[htb!]
\begin{center}
\resizebox{0.6\textwidth}{!}{%
\includegraphics{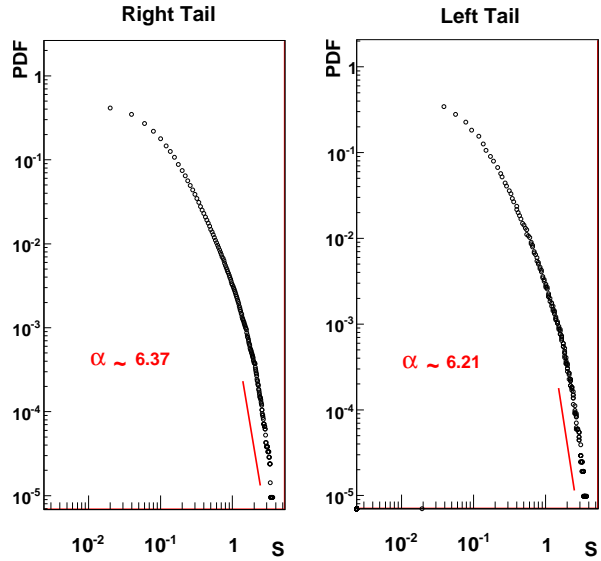}}
\caption{Returns PDF. a) Right tail b) Left tail. In total, after the cut off we have analyzed $479940$ events. Red lines are only sight guides. }
\label{pdf}
\end{center}
\end{figure}

\vspace*{-.9cm}
\noindent
Returns distribution $S_i$, for all of our generated samples are shown in 
figure \ref{Returns}a). Figures \ref{Returns}b) and  \ref{Returns}c) show the 
returns distribution in a log-log plot for right and left tails respectively. 
After applying the cut off to the sample, we construct the Probability Distribution 
Function (PDF) of the returns distribution; this is shown in figure \ref{pdf}. As it can be seen, both PDF tails decay as a power law with exponent $\alpha \sim 6$; Although it seems that there are more than one power law present along both curves.

\vspace*{-.7cm}
\subsection{Returns and Absolute Returns Autocorrelations}
\begin{figure}[htb!]
\begin{center}
\resizebox{0.6\textwidth}{!}{%
\includegraphics{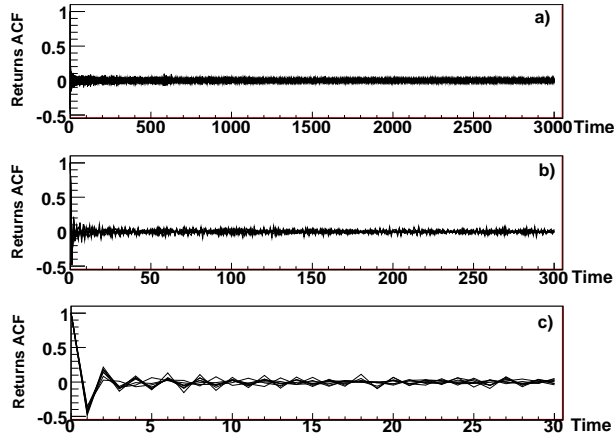}}
\caption{Returns ACF calculated for 8000 time steps and different time ranges.}
\label{acf}
\end{center}
\end{figure}

\begin{figure}[htb!]
\begin{center}
\resizebox{0.6\textwidth}{!}{%
\includegraphics{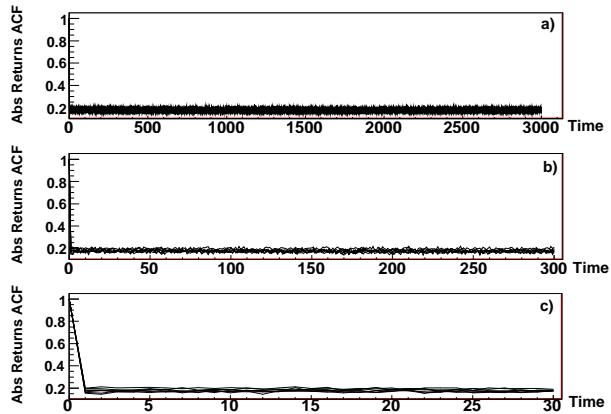}}
\caption{Absolute Returns ACF also calculated for 8000 time steps. Different time ranges are shown}
\label{aacf}
\end{center}
\end{figure}

\vspace*{-.9cm}
\noindent
Figures \ref{acf} and \ref{aacf} show the Auto Correlation Function (ACF) of 
returns and absolute returns for a few realizations of our experiments  respectively. It can be seen that returns ACF shows no memory, 
decaying to noise level almost immediately (in fact figure \ref{acf} looks 
similar to the ACF of a daily financial ACF time series). On the other hand, 
ACF of absolute returns decays quickly to a positive level of about 0.2 and stays fluctuating there, showing a very long range memory; all this in agreement with the stylized facts.

\vspace*{-.5cm}
\noindent
It is important to remark that although curve of figure \ref{acf} does not show correlations, a periodic pattern in each ACF there exists (although it was not shown here); also we have to mention that figure \ref{aacf} should look different wether the Absolute Returns ACF were calculated along the full $20000$  time steps, showing a slower decay more similar to that of financial time series. This behaviour was not showed here also.  
\vspace*{-.1cm}
\subsection{Volatility Distribution}
\begin{figure}[htb!]
\begin{center}
\resizebox{0.6\textwidth}{!}{%
\includegraphics{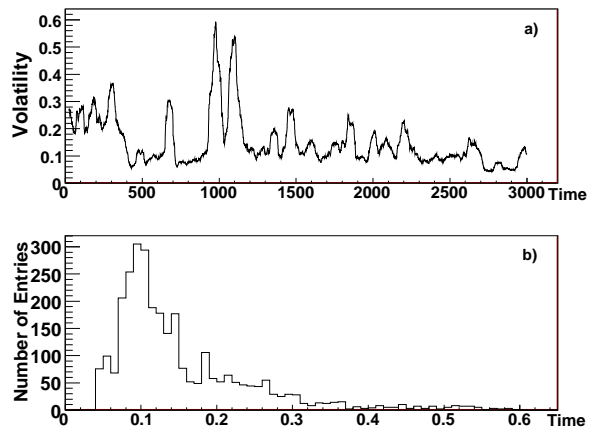}}
\caption{ a) Volatility for a typical generated random walk with a time
  window of 50 time steps. b) Its  distribution.}
\label{vola}
\end{center}
\end{figure}

\begin{figure}[htb!]
\begin{center}
\resizebox{0.6\textwidth}{!}{%
\includegraphics{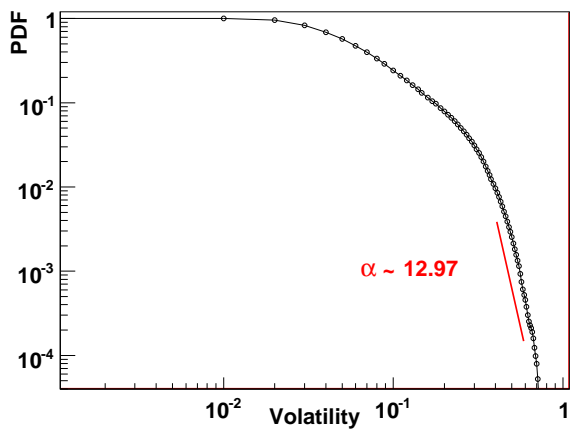}}
\caption{Volatility Probability Distribution Function for all the
  sample. It decays as a power law with exponent $\alpha  \sim 9$. }
\label{voladist}
\end{center}
\end{figure}

\vspace*{-.9cm}
\noindent
In this section, the volatility $V(t)$ is calculated \cite{calc_vola} by
averaging the absolute returns over a time window $T=n \Delta t$ as follows: 

\begin{equation}
V(t):= \frac{1}{n}\sum_{t'=t}^{t+n-1}|S(t')|,
\end{equation}
\noindent
Here we have set up  $\Delta t = $ 1 time lag and a window of 50 time steps. 
Figure \ref{vola}a) displays the volatility for the first 
$3000$ time steps of our observable for one of our generated random 
walks. Lower frame of same figure displays the corresponding volatility distribution. Figure \ref{voladist} shows that volatility PDF decays asymptotically as a 
power law with $\alpha \sim 13$, again in concordance with the stylized facts. Although also in this figure, as in the PDF of retuns case, more than a single power law  are present along the curve.

\noindent

\vspace*{-.3cm}

\vspace*{-.9cm}
\section{Discussion and Future Work}
\vspace*{-.9cm}
\noindent
The objective of this work is to generate in the simplest possible way 
a time series that displays the stylized facts that are well known to
the economic community. By means of a very simple geometrical method,
and using a bidimensional Cellular Automaton, in our case the Game of Life, we generate a synthetic time series with the properties of having returns distribution with fat tails, 
clustering volatility, practically no returns autocorrelations, long memory 
in ACF of absolute returns and volatility distribution decaying as a power law.
We believe that the use of our scheme to generate artificial financial time series
can be of help in understanding the underlying mechanisms that govern the formation
of the stylized facts arising in real markets. Finally, and about the future work that has to be made, is important to understand if there exists a multipower or a mix up of power laws in returns and volatility distributions as well as the patterns presented in the ACFs of the returns and absolute returns.

{\bf Acknowledgments\\}
\noindent
We appreciate very useful suggestions from S. Jim\'enez. We also thank to A. Robles from Market Activity Flow for its support and very useful discussions. This work has been supported by Conacyt-Mexico under Grant 44598. Plots and the Analyses have been performed using ROOT \cite{root}.

\end{document}